\documentclass[final,authoryear]{elsarticle}

\usepackage{amsmath}
\usepackage{amssymb}
\usepackage{graphics,graphicx}
\usepackage{epstopdf}
\usepackage{hyperref}
\usepackage{color}
\usepackage{caption}
\usepackage{subcaption}
\usepackage{natbib}

\makeatletter
\def\ps@pprintTitle{%
 \let\@oddhead\@empty
 \let\@evenhead\@empty
 \def\@oddfoot{}%
 \let\@evenfoot\@oddfoot}
\makeatother

\newcommand{\fref}[1]{Fig.~\ref{#1}}

\begin{document}

\begin{frontmatter}

\title{Mechanics of polymer brush based soft active materials-- theory and experiments}

\author[mech]{M. Manav}
\ead{manav@alumni.ubc.ca}

\author[blood]{P. Anilkumar}
%\ead{manav@alumni.ubc.ca}

\author[mech]{A. Srikantha Phani\corref{cor1}}
\ead{srikanth@mech.ubc.ca}

\cortext[cor1]{Corresponding author}

\address[mech]{Mechanical engineering, University of British Columbia, Vancouver, BC V6T 1Z4, Canada}
\address[blood]{Centre for Blood Research, University of British Columbia, Vancouver, BC V6T 1Z3, Canada}

\begin{abstract}
Brush-like structures emerge from stretching of long polymer chains, densely grafted on to the surface of an impermeable substrate.  They arise due to the competition between conformational entropic elasticity of  polymer chains and excluded volume interactions from the intra and interchain monomer repulsions. Recently, stimuli responsive polymer brush based soft materials have been developed to produce controllable and reversible large deformations of the host  substrate. To understand these systems, and improve their functional properties,  we study elastic stress distribution and surface stress-curvature relations of a neutral polymer brush grafted on to an elastic beam, made of a soft material. In the strongly stretched brush regime, we combine mean field theory from polymer physics with a continuum mechanics model and show that the residual stress variation  is a quartic function of distance from the grafting surface, with maximum stress occurring at the grafted surface. Idealizing the brush as a continuum elastic surface layer with residual stress, we derive a closed form expression for surface stress and the surface elasticity of the layer as a function of  brush parameters, such as graft density and molecular weight.   The  generalized continuum beam model accounts for the  Young-Laplace  and  Ogden-Steigman  curvature elasticity correction terms, and yields a surface stress-curvature relation, which contains existing relations in the literature as special cases.  Further, we report experiments on a thermoresponsive random copolymer brush, Poly(N- isopropylacrylamide)-co-Poly(N,N-Dimethylacrylamide) (PNIPAm-co-PDMA) brush, grafted on one side of a plasticized poly(vinyl chloride) (pPVC) thin film.  Estimated surface stress from measured curvature is on the order of $-10~N/m$, and it decreases gradually, and reversibly, with increasing ambient temperature from $15~^\circ{}C$ to $55~^\circ{}C$. 

\end{abstract}

\begin{keyword}
Polymer brush, Mean field theory, surface stress, surface elasticity, curvature elasticity
\end{keyword}

\end{frontmatter}

%%%%%%%%%%%%%%%%%%%%%%%%%%%%%%%%%%%%%%%%%%%%%
\section{Introduction}
\begin{figure}[!h]
	\center
	\includegraphics[width=4.6cm]{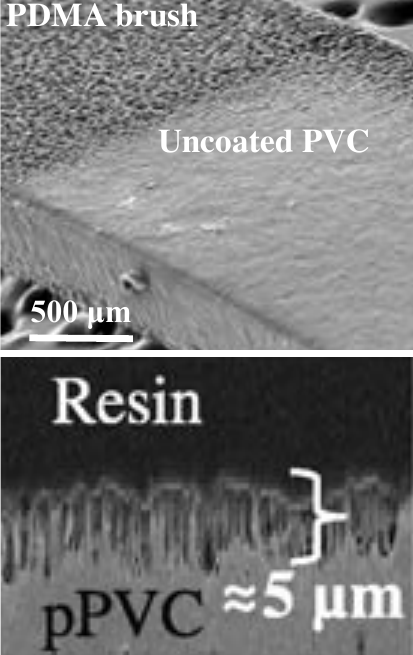}
	\includegraphics[width=6.1cm]{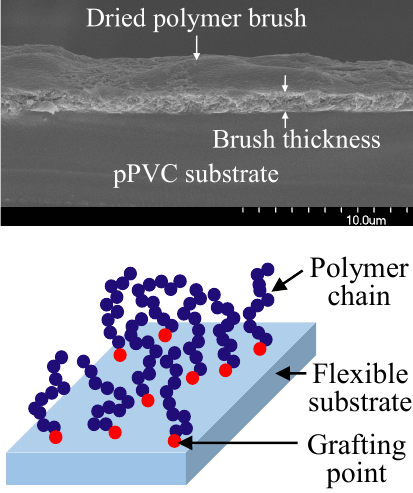}
	\caption{Scanning electron microscope (SEM) image of polymer brush coated on a plasticized poly(vinyl chloride) (pPVC) substrate~\citep{zou2011} (left), and an SEM image of a cross-section of a pPVC substrate with dried polymer brush along with an schematic of polymer chains grafted to the surface of a substrate (right).}
	\label{polymer_brush}
\end{figure}

Polymer brushes (PB)~\citep{milner91,edmondson2004} (see~\fref{polymer_brush}) are soft active materials (SAMs) that produce  reversible deformation~\citep{zou2011} in response to external stimuli, such as a change in temperature, pH, light, electromagnetic fields, among others. Other examples of SAMs include, stimuli-responsive gels~\citep{ahn2008,white2013}, electroactive polymers~\citep{scrosati93,wang2016}, liquid crystal elastomers~\citep{kupfer91,white2015} and shape memory polymers~\citep{lendlein2002,hager2015}.  Growing technological applications of these materials in the areas of drug delivery, self assembly, therapeutics, biomedical engineering, soft robotics etc.~\citep{hamley2003,bawa2009,albert2010,de2005,majidi2014} have led to considerable scientific interest in understanding their mechanical response~\citep{hong2008,utz08,wang2015}.   Stimuli response originating from \emph{surface} modification,  a unique feature of PB-SAMs, is advantageous due to minimal trade-off with other bulk material properties. Polymer brushes have been used as programmable material \citep{kelby2011} for  sensing and actuation~\citep{abu2006micro,klushin2014}, as microcantilever coatings in glucose sensing~\citep{chen2010}, selective metallic ion sensing~\citep{peng2017}, microcantilever actuation~\citep{zhou2006,zhou2008}, and  as macroscale bending stretching actuators~\citep{zou2011} exhibiting large elastic deformations.  A comprehensive overview of polymer brushes and their applications is available in~\citet{stuart2010} and~\citet{azzaroni2012}.

End-grafted long polymer chains adapt a stretched configuration in the presence of a good solvent\footnote{In a good  solvent, monomers have a high affinity for solvent molecules whereas in a poor solvent monomers minimize interaction with solvent molecules. In a $\theta$-solvent nearest neighbour monomer repulsions are absent.} at a sufficiently high grafting density at which neighbouring chains interact~\citep{alexander77,de80,milner91}. This brush-like structure (see~\fref{polymer_brush}), reminiscent of bristles on a tooth brush, is governed by a combination of strong entropic repulsion between monomers in a crowded monolayer, entropic stretching of polymer chains and constraints set by end grafting. The entropic spring force  strives to bring the two ends of a polymer chain together as a closed chain configuration has maximum entropy. Excluded volume effects among the monomers, present within  and neighbouring chains, tend to extend the chain. In a good solvent these repulsive interactions compete with entropic  elasticity and the chains extend away from the substrate, forming a brush structure. Significantly, these interactions make a brush behave like an elastic surface layer with residual surface stress which can be changed depending on the solvent quality. These stresses deform the underlying elastic substrate~\citep{zou2011,utz08}. Further, in the presence of an external stimulus  the interaction between monomers and the solvent is altered, such as through hydrogen bond breaking between polymer and solvent. This makes the solvent a poor solvent, leading to collapsed chain structure. In a stimuli-responsive polymer brush, residual surface stress as well as surface elasticity can be reversibly modified by switching  between stretched (good solvent) and collapsed (poor solvent) states. A  wide range of applications exist for stimuli sensitive polymer brushes, as detailed earlier.

Several theoretical approaches exist in polymer physics to describe the statistical mechanics of polymer brushes, see~\citet{halperin94}, \citet{netz03} and~\citet{binder12} for a review. The pioneering \emph{qualitative} scaling theories of~\citet{alexander77} and~\citet{de80} have been refined, and, in strong-stretching limit of polymer chains \emph{quantitative} mean-field theories have been proposed, whose predictions have been verified by molecular dynamic studies~\citep{netz03}. Scaling arguments establish a power law dependence of macroscopic properties of a brush on its molecular scale parameters. Typically the brush height ($H$) is related with number of monomers in a polymer chain ($N$), effective monomer size ($a$), and graft density\footnote{Number of polymer grafting points on unit area of the substrate} ($\rho_g$), according to the scaling relation: $H\sim Na\rho_g^{1/3}$. A remarkable feature of this scaling relation is that for sufficiently large $\rho_g$, the height of the brush $H$ is larger than the Flory radius\footnote{Quantifies the volume occupied by a single polymer chain with no neighbouring chain} ($R_F$), so that the polymer chains are strongly stretched to avoid high monomer density.  However, scaling theory assumes a \emph{step profile} for monomer density  and further assumes that all free ends of chains are at same height above the grafting surface. These restrictions are relaxed in mean filed theory. In a mean field theory~\citep{scheutjens79}, each polymer chain is considered  to be in a position-dependent mean field, which is equivalent to interaction with the surrounding monomers, and the mean field in turn is dependent on monomer density. Density profile is obtained numerically by utilizing the fact that the monomer density and the mean field at the minimum free energy configuration of the brush are self consistent~\citep{cosgrove87,milner90}. The mean field and scaling theories differ in the structure of the polymer brush, particularly in terms of chain end distribution throughout the height of the brush and monomer density profile. In the regime of strong stretching~\citep{netz98}, the partition function of a brush is dominated by the \emph{classical paths} of the chains, and other stochastic fluctuations about these paths can be ignored~\citep{semenov85}. This forms the basis for strong stretching theory (SST) which provides semianalytical and analytical results for the brush structure~\citep{milner88,skvortsov88,zhulina91}. In a moderately dense brush ($\rho_g a^2\lesssim 0.1$), binary interactions between monomers are dominant. Using only binary interactions between monomers in SST leads to the analytical prediction of parabolic monomer density profile among other properties of the brush~\citep{milner88,skvortsov88}. Parabolic profile has been observed in experiment~\citep{auroy92} and has also been confirmed by molecular dynamics simulation studies~\citep{murat89,grest93,dimitrov07} in the bulk of the brush. However deviations from parabolic profile is observed due to depletion layers near the grafted end, and a tail near the free end. Both the mean field theory and scaling theory yield the same scaling relation for the dependence of height on brush parameters (see~Table~\ref{height_comparison}), but their prediction of brush free energy is not the same. Table~\ref{freeEnergy_comparison} compares free-energy expressions used in three different approaches: global Flory argument for the entire brush (see \ref{brushHeight}), SST (local Flory argument), and scaling theory. The difference in scaling occurs because unlike scaling theory, mean field theory does not capture the excluded volume correlations that occur in the limit of strong excluded volume interactions. This also limits the applicability of mean field theory to brushes with weak excluded volume interactions ($v^2<<\rho_g a^8<<1$, where $v$ is binary interaction parameter)~\citep{milner88,kreer04}. Note that strong stretching does not necessarily require strong excluded volume interactions and can be brought about by other parameters such as high graft density.

\begin{table}[!h]
	\caption{\label{freeEnergy_comparison} Comparison of expressions for free energy of a brush in good solvent obtained from mean field Flory theory, SST and scaling theory. $k_B$ is Boltzmann constant and $T$ is temperature in Kelvin. Note that the scaling exponent of $a$ changes drastically between SST and scaling theory due to change in scaling exponent of  $v~(\sim a^3)$ and also of $\rho_g$.}
	\center
	\begin{tabular}{|l|l|}
		\hline
		  & Free energy in a good solvent (in $k_BT$ units) \\ \hline
		Flory argument & $\frac{9}{2}\left(\frac{1}{6}\right)^{2/3}v^{2/3}\rho_g^{10/6}a^{-2/3}N$ \\ \hline
		SST (local Flory argument) & $\frac{9}{10}\left(\frac{\pi^2}{4}\right)^{1/3}v^{2/3}\rho_g^{10/6}a^{-2/3}N$ \\ \hline
		Scaling theory & $\sim \frac{5}{2}v^{1/3}\rho_g^{11/6}a^{2/3}N$ \\ \hline
	\end{tabular}
\end{table}

In this work, stress in a polymer brush grafted to a rigid substrate is derived using mean field theory for brushes~\citep{milner88,zhulina91} in good and $\theta$- solvents. The remainder of this article is organized as follows.  A relation between the molecular scale parameters of a brush, and the surface stress and surface elasticity due to polymer brush is obtained in Section~\ref{SurfS_pb}, by treating polymer brush as a surface layer.  It will be shown from free energy comparisons that a polymer brush can produce large bending deformation. Then, a mechanics model for a thin flexible beam coated with a polymer brush layer on its top surface is developed in Section~\ref{mechanics_beam}, using virtual work principle and incorporating Young-Laplace and curvature elasticity corrections. We measure the curvature of a polymer brush coated beam in Section~\ref{exp} and  use the equations from Section~\ref{mechanics_beam} to estimate surface stress, ending with concluding remarks in Section~\ref{conclusion}.

%%%%%%%%%%%%%%%%%%%%%%%%%%%%%%%%%%%%%%%%%%%%%%
\section{Stress in a polymer brush using mean field theory}
\label{SurfS_pb}
\begin{figure}[!h]
\center
\includegraphics[width=10cm]{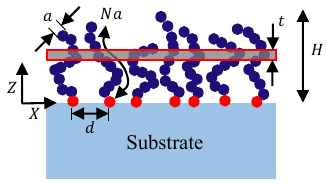}
\caption{A schematic showing side view of a planar polymer brush with height $H$. Each bead represents an effective monomer with size $a$. Contour length of a polymer chain is $Na$. Average distance between grafting points is inverse square root of graft density ($\langle d\rangle=\rho_g^{-1/2}$). A thin layer at heigh $Z$ is also shown.}
\label{brush_slit}
\end{figure} 
Consider the schematic of a planar polymer brush shown in~\fref{brush_slit}. Free energy of the brush is the sum of interaction free energy $F_{int}$ and chain stretching free energy $F_{el}$. Interaction free energy is a function of monomer number density $\phi(Z)$ and two functions are required to evaluate it: $g(\zeta)$, quantifying the number of chains per unit substrate area ending at height $\zeta$ ($\int_0^H g(\zeta)d\zeta=\rho_g$), and $E(Z,\zeta)$, local stretching at $Z$ in polymer chains with ends at height $\zeta$ ($Z\le\zeta\le H$), as stretching in a chain is not uniform along the length. Free energy can then be expressed as:
\begin{align}
F & =F_{int}+F_{el}\nonumber \\
&=\frac{1}{2}k_BT\int_0^H v\phi^2(Z) dZ+\frac{1}{6}k_BT\int_0^H w\phi^3(Z) dZ\nonumber\\
&+\frac{1}{2}k_BT\beta\int_0^H \int_{Z}^{H}g(\zeta)E(Z,\zeta)d\zeta  dZ,\quad \beta=\frac{3}{pa^2},\quad p=\frac{l_k}{a},
\label{eqs3a}
\end{align}
where $v$ and $w$, respectively, are the binary and ternary interaction parameters. $p$ is \emph{asymmetry} parameter, and $l_k$ is Kuhn length of polymer. $p=1$ for a flexible polymer chain and $p>1$ for a semiflexible chain.

Monomer number density $\phi(Z)$ is related to $g(\zeta)$ and $E(Z,\zeta)$ as follows~\citep{zhulina91}:
\begin{equation}
\phi(Z)=\displaystyle\int_Z^H \frac{g(\zeta)}{E(Z,\zeta)}d\zeta.
\label{eqs1a}
\end{equation}
with the normalization conditions:
\begin{equation}
\displaystyle\int_0^\zeta \frac{1}{E(Z,\zeta)}dZ=N, \quad \displaystyle\int_0^H \phi(Z) dZ
=\rho_g N.
\label{eqs1b}
\end{equation}
Minimization of free energy in~\eqref{eqs3a} with respect to $g(\zeta)$ and $E(Z,\zeta)$, under the constraints~\eqref{eqs1b} yields the equilibrium properties of the brush~\citep{zhulina91}. Considerable simplification arises under good solvent conditions for moderate graft density brushes (the second term containing  $w$ in~\eqref{eqs3a} is negligble) and $\theta$-solvent ($v=0$) conditions. In the following, we derive stress expressions for these two cases.

\subsection{Good solvent}
\label{stress_in_brush}
In brush regimes where binary interaction dominates, contribution from ternary interaction  to the free energy density ($w$ term in~\eqref{eqs3a}) can be ignored, leading to the following simplified expression:
\begin{align}
F & =F_{int}+F_{el}\nonumber \\
&\approx \frac{1}{2}k_BT\int_0^H v\phi^2(Z) dZ+\frac{1}{2}k_BT\beta\int_0^H \int_{Z}^{H}g(\zeta)E(Z,\zeta)d\zeta  dZ.
\label{eqs3f}
\end{align}
Minimization of free energy in this case yields analytical expressions for the unknown functions $\phi(Z)$, $g(\zeta)$ and $E(Z,\zeta)$, as well as $H$ \citep{milner88,zhulina91} as given below:
\begin{eqnarray}
\phi(Z)=\frac{\pi^2\beta}{8N^2v}\left(H^2-Z^2\right),\label{eqs3b}\\
H=\left(\frac{12}{\pi^2}\right)^{1/3}\frac{v^{1/3}\rho_g^{1/3}N}{\beta^{1/3}}, \label{eqs3c} \\
g(\zeta)=\gamma\zeta\sqrt{H^2-\zeta^2}, \quad \gamma=\frac{\pi^2\beta}{4N^3v}, \label{eqs3d} \\
E(Z,\zeta)=\frac{\pi}{2N}\sqrt{\zeta^2-Z^2}. \label{eqs3e}
\end{eqnarray}

Free energy distribution obtained from SST for a planar brush is also nonuniform and it shows a variation through height of the brush. By calculating change in free energy density in a brush due to a uniform horizontal strain applied to it, we can obtain stress distribution within the brush.

In order to calculate lateral stress variation in height direction, we consider an infinitesimally thin rectangular layer of brush (\fref{brush_slit}) of unit horizontal area and of small thickness $t$ at height $Z$ above the grafted surface. Free energy of this infinitesimal layer is given by:
\begin{align}
\Delta F & =\Delta F_{int}+\Delta F_{el}\nonumber \\
&=\left[\left(\frac{1}{2}v\phi^2(Z)+\frac{1}{2}\beta\int_{Z}^{H}g(\zeta)E(Z,\zeta)d\zeta  \right)k_BT\right]t=f(Z)t,
\label{eqs3}
\end{align}
where $f(Z)= \Delta F/t$ is free energy density at height $Z$. Brush has residual stress $\sigma_{IJ},~I,J=X,~Y,~Z$. For an infinitesimal strain $\epsilon_{IJ}$ experienced by the thin layer when an infinitesimal strain $\epsilon_{\textsc{xx}}$ is applied to the substrate,
\begin{equation}
\Delta(f(Z)V_0)=\sigma_{IJ}\epsilon_{IJ}V_0+\mu \Delta N,
\label{eqs4}
\end{equation}
where $V_0$ is the initial volume of the thin layer; $\mu$ is chemical potential of solvent in the layer and $\Delta N$ is the change in the number of solvent molecules in the thin layer due to the application of strain.  We further assume transverse-isotropy of  the polymer brush layer and ignore Poisson coupling. 

Assuming plain strain condition in $Y$ direction, \eqref{eqs4} converts to the following:
\begin{equation}
\frac{\Delta(f(Z)V_0)}{V_0}=\sigma_{\textsc{xx}}\epsilon_{\textsc{xx}}+\frac{\mu\Delta N}{V_0}.
\label{eqs5}
\end{equation}
As the polymer molecules and the solvent are incompressible,
\begin{align}
\frac{\Delta V}{V_0}=\epsilon_{\textsc{xx}}+\epsilon_{\textsc{zz}}=\frac{\Omega\Delta N}{V_0}, \\
\frac{\Delta N}{V_0}=\frac{\epsilon_{\textsc{xx}}+\epsilon_{\textsc{zz}}}{\Omega},
\label{eqs6}
\end{align}
where $\Omega$ is volume of a solvent molecule. Substituting the above in \eqref{eqs5} and diving by $\epsilon_{\textsc{xx}}$ yields,
\begin{equation}
\frac{1}{V_0}\frac{\Delta(f(Z)V_0)}{\epsilon_{\textsc{xx}}}=\left(\sigma_{\textsc{xx}}+\frac{\mu}{\Omega}\right)+\frac{\mu}{\Omega}\frac{\epsilon_{\textsc{zz}}}{\epsilon_{\textsc{xx}}}.
\label{eqs7}
\end{equation}
Expanding the left hand side term in the above and for $\epsilon_{\textsc{xx}}$ approaching $0$, we get,
\begin{align}
\frac{\partial f(Z)}{\partial\epsilon_{\textsc{xx}}}+f(Z)\left(1+\frac{\partial\epsilon_{\textsc{zz}}}{\partial\epsilon_{\textsc{xx}}}\right)=\left(\sigma_{\textsc{xx}}+\frac{\mu}{\Omega}\right)+\frac{\mu}{\Omega}\frac{\partial\epsilon_{\textsc{zz}}}{\partial\epsilon_{\textsc{xx}}}.
\label{eqs8}
\end{align}
As we are interested only in equilibrium conditions, $\mu=0$. Substituting this in the above, we obtain the expression for residual stress in $X$ direction.
\begin{equation}
\sigma_{\textsc{xx}}=\frac{\partial f(Z)}{\partial\epsilon_{\textsc{xx}}}+f(Z)\left(1+\frac{\partial\epsilon_{\textsc{zz}}}{\partial\epsilon_{\textsc{xx}}}\right).
\label{eqs9}
\end{equation}
Note that the above equation is independent of the free energy density expression and can be used for SST with interactions of order higher than binary.

To evaluate the above expression, we need to find $\frac{\partial f(Z)}{\partial\epsilon_{\textsc{xx}}}$ and $\frac{\partial\epsilon_{\textsc{zz}}}{\partial\epsilon_{\textsc{xx}}}$. To evaluate $\frac{\partial\epsilon_{\textsc{zz}}}{\partial\epsilon_{\textsc{xx}}}$, we make use of the fact that the monomers in the layer of thickness $t$ in the initial configuration remain in the layer after strain $\epsilon_{\textsc{xx}}$ is applied on the substrate, though the layer displaces by $u_3$ in $Z$-direction as the brush reaches a new equilibrium. So, volume occupied by monomers remains unchanged but monomer density $(\phi(Z))$ changes due to a change in volume of the layer. Hence,
\begin{eqnarray}
\Delta(\phi(Z) V_0)=0,\nonumber \\
\frac{\partial \phi(Z)}{\partial\epsilon_{\textsc{xx}}}+\phi(Z)\left(1+\frac{\partial\epsilon_{\textsc{zz}}}{\partial\epsilon_{\textsc{xx}}}\right)=0.
\label{eqs10}
\end{eqnarray}
Using \eqref{eqs3b} in the above, we obtain:
\begin{eqnarray}
\frac{\partial \phi(Z)}{\partial\epsilon_{\textsc{xx}}}=\frac{\pi^2\beta}{4N^2v}\left(H\frac{\partial H}{\partial\epsilon_{\textsc{xx}}}-Z\frac{\partial u_3}{\partial\epsilon_{\textsc{xx}}}\right).
\label{eqs11}
\end{eqnarray}
We also notice that
\begin{align}
\frac{\partial\epsilon_{\textsc{zz}}}{\partial\epsilon_{\textsc{xx}}}=&\frac{\partial}{\partial\epsilon_{\textsc{xx}}}\left(\frac{\partial u_3}{\partial Z}\right)=\frac{\partial}{\partial Z}\left(\frac{\partial u_3}{\partial \epsilon_{\textsc{xx}}}\right).
\label{eqs12}
\end{align}
Using~\eqref{eqs11} and~\eqref{eqs12} in~\eqref{eqs10}:
\begin{eqnarray}
\frac{\pi^2\beta}{4N^2v}\left(H\frac{\partial H}{\partial\epsilon_{\textsc{xx}}}-Z\frac{\partial u_3}{\partial\epsilon_{\textsc{xx}}}\right)+ \phi(Z)\left(1+\frac{\partial}{\partial Z}\left(\frac{\partial u_3}{\partial \epsilon_{\textsc{xx}}}\right)\right)=0.
\label{eqs13}
\end{eqnarray}
We can use \eqref{eqs3c} to evaluate $\frac{\partial H}{\partial\epsilon_{\textsc{xx}}}$. Note that $\epsilon_{\textsc{xx}}$ changes $H$ by changing graft density $\rho_g$. Due to a change in surface area ($A^{deformed}=A(1+\epsilon_{\textsc{xx}})$), $\rho_g^{deformed}\approx\rho_g(1-\epsilon_{\textsc{xx}})$. Hence,
\begin{eqnarray}
\frac{\partial\rho_g}{\partial \epsilon_{\textsc{xx}}}=-\rho_g, \nonumber\\
\frac{\partial H}{\partial\epsilon_{\textsc{xx}}}=-\frac{1}{3}H.
\label{eqs15}
\end{eqnarray}
Substituting the above in \eqref{eqs13} yields:
\begin{eqnarray}
\frac{\pi^2\beta}{4N^2v}\left(-\frac{1}{3}H^2-Z\frac{\partial u_3}{\partial\epsilon_{\textsc{xx}}}\right)+ \phi(Z)\left(1+\frac{\partial}{\partial Z}\left(\frac{\partial u_3}{\partial \epsilon_{\textsc{xx}}}\right)\right)=0.
\label{eqs16}
\end{eqnarray}
As $\phi(Z)$ is a quadratic function in $Z$ (see \eqref{eqs3b}), we assume $\frac{\partial u_3}{\partial\epsilon_{\textsc{xx}}}=\sum\limits_{i=0} {a_nZ^n}$. Substituting this in the above equation and recognizing that $\left(\frac{\partial u_3}{\partial\epsilon_{\textsc{xx}}}\right)_{Z=H}=\frac{\partial H}{\partial\epsilon_{\textsc{xx}}}=-\frac{1}{3}H$, we obtain,
\begin{equation}
\frac{\partial u_3}{\partial\epsilon_{\textsc{xx}}}=-\frac{1}{3}Z.
\label{eqs17}
\end{equation}
Substituting the above in \eqref{eqs11} and \eqref{eqs12} gives:
\begin{eqnarray}
\frac{\partial\phi(Z)}{\partial\epsilon_{\textsc{xx}}}=-\frac{2}{3}\phi(Z), \label{eqs18} \\
\frac{\partial\epsilon_{\textsc{zz}}}{\partial\epsilon_{\textsc{xx}}}=-\frac{1}{3}.
\label{eqs19}
\end{eqnarray}

Now, we evaluate $\frac{\partial f(Z)}{\partial\epsilon_{\textsc{xx}}}$ using the expression in \eqref{eqs3}.
\begin{align}
\frac{\partial f(Z)}{\partial\epsilon_{\textsc{xx}}}=& v\phi(Z)\frac{\partial\phi(Z)}{\partial\epsilon_{\textsc{xx}}}k_BT\nonumber \\
&+\frac{1}{2}k_BT\beta\left(g(H)E(Z,H)\frac{\partial H}{\partial\epsilon_{\textsc{xx}}}-g(Z)E(Z,Z)\frac{\partial u_3}{\partial\epsilon_{\textsc{xx}}}\right)\nonumber \\
&+\frac{1}{2}k_BT\beta\left(\int_Z^H g(\zeta)\frac{\partial E(Z,\zeta)}{\partial\epsilon_{\textsc{xx}}}d\zeta+\int_Z^H \frac{\partial g(\zeta)}{\partial\epsilon_{\textsc{xx}}}E(Z,\zeta) d\zeta \right).
\label{eqs20}
\end{align}
From \eqref{eqs3d} and \eqref{eqs3e}, $g(H)=0$ and $E(Z,Z)=0$, and
\begin{eqnarray}
\frac{\partial g(\zeta)}{\partial\epsilon_{\textsc{xx}}}=\frac{\gamma \zeta H}{\sqrt{H^2-\zeta^2}}\frac{\partial H}{\partial\epsilon_{\textsc{xx}}}=-\frac{\gamma \zeta H^2}{3\sqrt{H^2-\zeta^2}}, \nonumber\\
\frac{\partial E(Z,\zeta)}{\partial\epsilon_{\textsc{xx}}}=-\frac{\pi}{2N}\frac{Z}{\sqrt{\zeta^2-Z^2}}\frac{\partial u_3}{\partial\epsilon_{\textsc{xx}}}= \frac{\pi}{6N}\frac{Z^2}{\sqrt{\zeta^2-Z^2}}.
\label{eqs22}
\end{eqnarray}
$g(H)=0$ means that the density of free ends at $H$ is zero. $E(Z,Z)=0$ means that the stretch in a polymer chain at its free end is zero.

The expression for $\sigma_{\textsc{xx}}$ in \eqref{eqs9} transforms to the following on substituting the relations in \eqref{eqs18}, \eqref{eqs19} and \eqref{eqs20}.
\begin{align}
\sigma_{\textsc{xx}}=&-\frac{1}{3}v\phi^2(Z)k_BT \nonumber\\
&+\frac{1}{2}k_BT\beta\int_Z^H g(\zeta)E(Z,\zeta)d\zeta \nonumber\\
&+\frac{1}{2}k_BT\beta\left(\int_Z^H g(\zeta)\frac{\partial E(Z,\zeta)}{\partial\epsilon_{\textsc{xx}}}d\zeta+\int_Z^H \frac{\partial g(\zeta)}{\partial\epsilon_{\textsc{xx}}}E(Z,\zeta) d\zeta \right).
\label{eqs23}
\end{align}
In the above expression, the first term involving $v$ results from nonbonded interaction. The remaining terms result from polymer chain stretching. Substituting \eqref{eqs3d}, \eqref{eqs3e} and \eqref{eqs22} in the above and carrying out the integration yields:
\begin{align}
&\sigma_{\textsc{xx}}=\underbrace{-\frac{\pi^4\beta^2 k_BT}{192vN^4}\left(H^2-Z^2\right)^2}_\text{Nonbonded interaction}\underbrace{-\frac{\pi^4\beta^2 k_BT}{384vN^4}\left(H^2-Z^2\right)^2}_\text{Chain stretching}\nonumber\\
&=-\frac{\pi^4\beta^2 k_BT}{128vN^4}\left(H^2-Z^2\right)^2=-\frac{9}{8}\left(\frac{\pi^2}{12}\right)^{2/3}v^{1/3}\rho_g^{4/3}\beta^{2/3}\left(1-\left(\frac{Z}{H}\right)^2\right)^2k_BT.
\label{eqs24}
\end{align}
We observe that the nonbonded interactions contribute \emph{twice} as much as the chain stretching in the expression for stress in a brush layer. Stress shows \emph{strong} dependence on graft density. Its dependence on chain length is through brush height $H$. Also, stress goes to zero at $Z=H$ smoothly (with zero slope and curvature) compared to monomer density $\phi(Z)$ (see \eqref{eqs3b}).

It should be observed that since a plane shear strain ($\epsilon_{\textsc{xy}}$) on the surface of a substrate does not cause any change in surface area and consequently in graft density and free energy of a brush, plane residual shear stress ($\sigma_{\textsc{xy}}$) and the associated shear modulus of the brush equal zero.

%%%%%%%%%%%%%%%%%%%%%%%%%%%%%%%%%%%
\subsection{$\theta$-solvent}
In a $\theta$-solvent $v=0$, and monomer-monomer interaction is governed by ternary interaction parameter $w$. Free energy density in a brush at a height $Z$ is given by \citep{zhulina91}:
\begin{equation}
f(Z)=\left(\frac{1}{6}w\phi^3(Z)+\frac{1}{2}\beta\int_{Z}^{H}g(\zeta)E(Z,\zeta)d\zeta \right)k_BT,
\label{eqs24a}
\end{equation}
and 
\begin{eqnarray}
\phi(Z)=\frac{\pi}{2N}\sqrt{\frac{\beta}{w}}\sqrt{H^2-Z^2},\label{eqs24b}\\
H=\frac{2}{\pi}\left(\frac{4w}{\beta}\right)^{1/4}\rho_g^{1/2}N, \label{eqs24c} \\
g(\zeta)=\frac{\pi^2}{4N^2}\sqrt{\frac{\beta}{w}}\zeta, \label{eqs24d} \\
E(Z,\zeta)=\frac{\pi}{2N}\sqrt{\zeta^2-Z^2}. \label{eqs24e}
\end{eqnarray}
Notice that monomer density  profile is not parabolic. Also, height of the brush varies as square root of graft density. The number density of chains with end at height $\zeta$ is an increasing function and it assumes the maximum value at height $H$. The local stretch in a chain remains the same as in a chain in a good solvent. Following the same method as in Section~\ref{stress_in_brush}, stress distribution in brush can be shown as:
\begin{align}
\sigma_{\textsc{xx}}=\frac{4^{3/4}}{3}\beta^{3/4}w^{1/4}\rho_g^{3/2}\left(1-\left(\frac{Z}{H} \right)^2 \right)^{3/2}k_BT, \label{eqs24f}
\end{align}
wherein nonbonded interaction and chain stretching contribute \emph{equally} in the expression for stress. Note that stress variation is \emph{no longer quartic} with respect to the distance from the grafting surface and dependence of stress on graft density becomes stronger. Dependence of stress on chain length is again through the brush height $H$.

%%%%%%%%%%%%%%%%%%%%%%%%%%%%%%%%%%%
\subsection{Surface stress and surface modulus}
\label{PB_surf}
The surface stress due to the brush layer is obtained by summation of stress in all the layers in a brush. In a good solvent,
\begin{equation}
\tau_s=\int_0^H \sigma_{\textsc{xx}} dZ=-\left(\frac{3}{5}\left(\frac{\pi^2}{12}\right)^{1/3}v^{2/3}\rho_g^{5/3}\beta^{1/3}N\right)k_BT.
\label{eqs2}
\end{equation}
We observe that $\tau_s$ has the same value for normal surface stress in any direction along the surface. Notice that $\tau_s$ has stronger dependence on graft density compared to number of monomers in a polymer chain.

To obtain surface elastic modulus, we need to determine the change in surface stress as a brush coated substrate is stretched. Stretching of substrate causes change in graft density, which leads to a change in surface stress. Change in graft density due to small stretching of a substrate is given by:
\begin{equation}
\rho_g^{deformed}=\rho_g(1-(\epsilon_{\textsc{xx}}+\epsilon_{YY})).
\end{equation}
Surface elastic modulus is obtained using the following relation:
\begin{equation}
E_s=\frac{\partial \tau_s}{\partial \epsilon_{\alpha\alpha}}=\left(\left(\frac{\pi^2}{12}\right)^{1/3}v^{2/3}\rho_g^{5/3}\beta^{1/3}N\right)k_BT, \quad \alpha=X,~Y.
\label{eqs2a}
\end{equation}
We observe that the scaling of the surface stress and the surface modulus are the same. Their magnitudes are also of the same order.

We can repeat the above calculation for a $\theta$-solvent to obtain:
\begin{eqnarray}
\tau_s=-\left(\frac{1}{2\pi}w^{1/2}\rho_g^2\beta^{1/2}N\right)k_BT,\label{eqs2b}\\
E_s=\left(\frac{1}{\pi}w^{1/2}\rho_g^2\beta^{1/2}N\right)k_BT.
\label{eqs2b1}
\end{eqnarray}

Table~\ref{stress_comparison} compares surface stress expressions obtained from different theories. Note that difference in scaling of stress in a good solvent between SST and scaling theory originates from difference in scaling of free energy predicted by the two theories (see Table~\ref{freeEnergy_comparison}), even though brush height predictions agree (see Table~\ref{height_comparison}). The difference in scaling originates from the fact that mean field theory and SST do not take into account the excluded volume correlations that occur in the case of strong excluded volume interactions. Also, for $\theta$-solvent, expression from scaling theory has an addition term coming from the term in the virial expansion of mean field free energy of mixing proportional to $\phi$ \citep{utz08}. For large $N$, this term is typically ignored in mean field theory and SST.
\begin{table}[!h]
	\caption{\label{stress_comparison} Comparison of expressions for surface stress (in $k_BT$ units) of a brush in good and $\theta$ solvents obtained from mean field Flory theory, SST and scaling theory \citep{utz08}. Note that in the expression from SST, $p=1$ and hence $\beta=3/a^2$.}
	\center
	\begin{tabular}{|l|l|l|}
		\hline
		 & Good solvent & $\theta$ solvent	 \\ \hline
		Flory argument & $-3\left(\frac{1}{6}\right)^{2/3}v^{2/3}\rho_g^{10/6}a^{-2/3}N$ & $-w^{1/2}\rho_g^2a^{-1}N$ \\ \hline
		SST (this work) & $-\frac{3}{5}\left(\frac{\pi^2}{4}\right)^{1/3}v^{2/3}\rho_g^{10/6}a^{-2/3}N$ & $-\frac{\sqrt{3}}{2\pi}w^{1/2}\rho_g^2a^{-1}N$ \\ \hline
		Scaling theory & $\sim -\frac{1}{3}v^{1/3}\rho_g^{11/6}a^{2/3}N$ & $
		\sim -\frac{1}{2}\left(\rho_g+w\rho_g^2a^{-4}N\right)$ \\ \hline
	\end{tabular}
\end{table}

%%%%%%%%%%%%%%%%%%%%%%%%%%%%%%%%%%%%%%%%%%%%%%
\subsection{Energetics of bending}
\label{EnergyComparison}
We now qualitatively compare the change in free energy of a polymer brush due to a strain in substrate with the strain energy of bending of an Euler beam and obtain Young's modulus of the beam which will bend substantially due to a brush layer. This comparison is meaningful only when the brush remains planar, that is the length scale associated with the substrate curvature is large compared to brush height.

For bending of an Euler beam of thickness $h$ due to brush grafted to its top surface,
\begin{equation}
|\tau_s\epsilon_s |\sim\frac{1}{2}\frac{Eh^3}{12}\kappa^2,
\label{eqs2c}
\end{equation}
where the term on the left is the change in the free energy of brush in a unit area because of strain due to bending in the substrate and the term on the right is bending strain energy per unit length of a beam with unit width. $\epsilon_s$ is strain at the top surface of the beam due to bending and $\kappa$ is curvature of bending. $E$ is Young's modulus of substrate. On neglecting mid plane stretching in the beam, $\epsilon_s=\kappa h/2$. Substituting this in the above relation yields:
\begin{equation}
E\sim\frac{12|\tau_s|}{h(\kappa h)}.
\label{eqs2d}
\end{equation}
Assuming that $\epsilon_s \sim 10^{-3}$, $\kappa h\sim 10^{-3}$. In this case, deflection of the tip of a cantilever beam of length $50h$ is $\sim h$. Now, we need to obtain $\tau_s$ to be able to estimate Young's modulus of a beam that will show this level of strain and deflection due to bending.

We use \eqref{eqs2} to estimate surface stress due to a brush. Excluded volume parameter $v=a^3(1-2\chi)$, where $\chi$ is Flory-Huggins interaction parameter, which characterizes interaction between solvent molecules and monomers. In a good solvent, $v\approx a^3$ as $\chi$ approaches $0$ \citep{de79}. We assume graft density $\rho_g\sim 0.1/a^2$. Note that $max(\rho_g a^2)\le1$. For brushes with high molecular weight polymer chains (fabricated using ATRP method), $N\sim 10^4$. Taking $a=1~nm$ and $p=1$  (flexible polymer), and at room temperature, $\tau_s^0\sim -1~N/m$ and $E_s\sim1~N/m$. Height of the brush is $\sim 1~\mu m$. For a beam with $100~\mu m$ thickness, bending is considerable if $E\sim 100~MPa$. Also, in this case radius of curvature is $\sim 10~cm$. Since the radius of curvature is much larger than the height of the brush, the brush is planar.

In summary, energetics of bending of a beam due to polymer brush suggests that polymer brush can produce large deformation in thin beams of materials with low elastic modulus.

%%%%%%%%%%%%%%%%%%%%%%%%%%%%%%%%%%%%%%%%%%%%%%

\section{Large deflection of a beam with a polymer brush layer}
\label{mechanics_beam}
\begin{figure}[!h]
	\center
	\includegraphics[width=8cm]{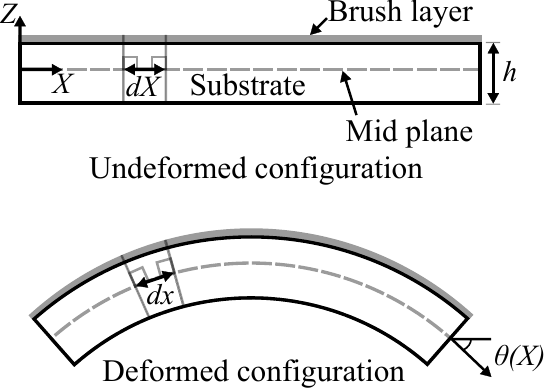}
	\caption{Schematic of an elastic beam with surface stress due to polymer brush layer in undeformed (top) and deformed states (bottom) of the substrate.}
	\label{beam_surfS}
\end{figure}

In this section, we develop a mechanics model for a thin flexible beam (Young's modulus $E$, Poisson's ratio $\nu$) with an elastic surface layer of nonzero thickness (elastic modulus $E_s$, Steigmann-Ogden constant $C$) with residual stress due to polymer brush at its top surface, using the principle of virtual work. Surface stress in the undeformed state of the substrate is $\tau_s^0$, and in the deformed beam, it is $\tau_s$. \fref{beam_surfS} shows the configuration with the undeformed substrate along with the deformed configuration. The model allows large deformation of the beam, however, strain should remain small (valid for thin beams). The absence of shear stress lets us assume that plane sections remain plane and perpendicular to the centreline of the beam.

Bending of a classical Euler-Bernoulli beam is governed by:
\begin{equation}
EI\frac{d\theta}{dx}-M=0,
\end{equation}
where $I$ is the second moment of area of cross-section of the beam and $M$ is bending moment at the cross-section. In the presence of a surface layer, effective Young's modulus is modified \citep{zhu09,wang10,liu2011,chhapadia2011}. Assuming identical surface layer at the top and bottom surfaces of a beam and accounting for mechanical equilibrium of surface led to inclusion of Young-Laplace correction term \citep{liu2011}. Accounting for nonzero thickness of surface necessitated introduction of curvature elasticity of surface \citep{chhapadia2011} in the effective modulus. Below, we review the conditions when the corrections become considerable and then formulate the governing equation for a beam with surface layer at its top surface by including the Young-Laplace and curvature elasticity corrections.

For very thin substrates, mechanical equilibrium of the surface layer introduces correction in stress in the substrate through Young-Laplace relation. For a rectangular cross section with thickness $h$, the correction is considerable if $\tau_s\sim\frac{Eh}{\nu}$. From our experiment, $\tau_s\sim -10~N/m$. Hence, correction due to Young-Laplace relation is considerable if $Eh\sim1~N/m$ ($\nu\sim10^{-1}$). For $E\sim10^6~Pa$, correction due to Young-Laplace relation is considerable for a micron thick beam.

Furthermore, the thickness of a polymer brush layer is $\sim 1~\mu m$ \citep{zou2011}. Hence, for very thin substrates, the surface layer can not be assumed to be of zero thickness and curvature dependence of surface elasticity \citep{steigmann99,chhapadia2011} may introduce considerable correction to effective modulus. For a rectangular beam with thickness $h$, effective elastic modulus \citep{chhapadia2011} is given as:
\begin{equation}
E_{eff}=E\left(1+\frac{3E_s}{Eh}+\frac{12C}{Eh^3} \right).
\label{eq0}
\end{equation}
The second term in the above relates to the elasticity of the surface layer. The ratio $E_s/E$ is a material length scale~\citep{miller2000}. The third term has its origin in curvature dependence of surface energy. Also, the effective thickness of a surface, $h_s$, has been defined as $h_s=\sqrt{CE_s^{-1}}$~\citep{chhapadia2011}. So, for a given thickness of a surface layer, $C=E_s h_s^2$. By substituting this relation in \eqref{eq0}, we can conclude that the correction to effective modulus ($E_{eff}$) due to Young's modulus of the surface and the curvature elasticity are of the same order if $h_s\sim h$. So for a polymer brush layer, the two corrections will be of similar order when the beam is a micron thick.

Now, we develop governing equations for deformation of the beam, shown in Fig.~\ref{beam_surfS}. Lagrangian strain in the beam is given by
\begin{equation}
\epsilon _{XX} (X,Z)=(\lambda(X)-1)-\frac{d\theta(X)}{dX}Z,\quad \lambda(X)=\frac{dx}{dX},
\label{eq1} 
\end{equation}
where $Z$ is the distance of a point from the centreline of the beam. We assume that the deformation of the beam does not change this distance. $\lambda(X)$ is mid plane stretch.

The axial stress (second Piola-Kirchhoff stress) in the beam is given by:
\begin{equation}
\sigma_{\textsc{xx}}(X,Z)=\bar{E}\epsilon_{\textsc{xx}}+\frac{\nu}{1-\nu}\sigma _{ZZ},
\label{eq2}
\end{equation}
where $\bar{E}=E/(1-\nu^2)$ (plane stress modulus). Normal stress in the bulk just underneath the surface layer in the direction normal to the centreline of the beam ($\sigma_{\textsc{zz}}^+$) can be obtained using Young-Laplace theory~\citep{liu2011}.
\begin{eqnarray}
\sigma_{\textsc{zz}}^+(X)=\lambda\frac{d\theta}{dX}\tau_s\approx\frac{d\theta}{dX}\tau_s, \label{eq3}
\end{eqnarray}
where
\begin{eqnarray}
\tau_s (X)=\tau_s^0+E_s\left ((\lambda-1)-\frac{d\theta}{dX}\frac{h}{2}\right ).
\label{eq4}
\end{eqnarray}
Also, since there is no brush on the bottom face of the beam $\sigma_{\textsc{zz}}^-(X)=0$. Assuming that $\sigma_{\textsc{zz}}$ varies linearly in $Z$ direction~\citep{lu2006}, it is given by the following expression:
\begin{equation}
\sigma_{\textsc{zz}}(X,Z)=\frac{1}{2}\frac{d\theta}{dX}\tau_s+\frac{Z}{h}\frac{d\theta}{dX}\tau_s.
\label{eq5}
\end{equation}
Substituting the above in \eqref{eq2} yields the expression for stress in the bulk.
\begin{equation}
\sigma_{\textsc{xx}}=\bar{E}\epsilon_{\textsc{xx}}+\frac{\nu}{1-\nu}\left( \frac{1}{2}\frac{d\theta}{dX}\tau_s+\frac{Z}{h}\frac{d\theta}{dX}\tau_s \right).
\label{eq6}
\end{equation}
Now, to obtain governing equation, we utilize the principle of virtual work. In equilibrium configuration,
\begin{equation}
\delta W_{int}+\delta W_{ext}=0,
\label{eq7}
\end{equation}
where $\delta W_{int}$ and $\delta W_{ext}$ are internal and external virtual works respectively. As there is no external force in our case $\delta W_{ext}=0$. Hence, (\ref{eq7}) reduces to $\delta W_{int}=0$. Internal virtual work has contribution from the bulk and the surface ($\delta W_{int}=\delta W_{bulk}+\delta W_{surf}$). The two contributions are given by:
\begin{eqnarray}
\delta W_{bulk}=\int_0^L {\left [\int_{-h/2}^{h/2} {\sigma_{\textsc{xx}}\delta \epsilon_{\textsc{xx}}{\rm d}Z} \right ] {\rm d}X}, \label{eq9} \\
\delta W_{surf}=\int_0^L {\left(\tau_s\left[\delta\epsilon_{\textsc{xx}}\right]_{Z=\frac{h}{2}} +C\frac{d\theta}{dX}\delta\left(\frac{d\theta}{dX}\right)\right){\rm d}X}, \label{eq9a}
\end{eqnarray}
where $C\frac{d\theta}{dX}$ is surface moment stress (curvature $\kappa=\frac{d\tilde{\theta}(x)}{dx}\approx\frac{d\theta}{dX}$ for small strain, where $\tilde{\theta}(x)=\theta(X(x))$)~\citep{steigmann99,chhapadia2011}. Note that we can use the linearized form of the surface moment stress~\citep{chhapadia2011}, because in our coated beam system, strain is small even though rotation may not. From (\ref{eq1}), variation in strain, $\delta \epsilon _{XX}$, can be written as:
\begin{equation}
\delta \epsilon _{XX}=\delta\lambda-Z\delta\left(\frac{d\theta}{dX}\right).
\label{eq10}
\end{equation}
On substituting the expressions corresponding to $\sigma_{\textsc{xx}}$ (\ref{eq6}) and $\delta \epsilon _{XX}$ (\ref{eq10}) in (\ref{eq9}), and integrating with respect to $Z$, the expression becomes
\begin{align}
\delta W_{bulk}=&\int_0^L {\left [h\left ( \bar{E}(\lambda-1)+ \frac{\nu}{1-\nu}\frac{\tau_s}{2}\frac{d\theta}{dX} \right )\delta \lambda\right ] {\rm d}X}+ \nonumber \\
&\int_0^L {\left [\frac{h^3}{12}\left (\bar{E}\frac{d\theta}{dX}-\frac{\nu}{1-\nu}\frac{\tau_s}{h}\frac{d\theta}{dX}  \right )\delta \left ( \frac{d\theta}{dX} \right ) \right ] {\rm d}X}.
\label{eq12}
\end{align}

The expression for the virtual work contribution from the surface layer, $\delta W_{surf}$ is given by:
\begin{eqnarray}
\delta W_{surf}=\int_0^L {\left(\tau_s\delta\lambda-\tau_s\frac{h}{2}\delta\left(\frac{d\theta}{dX}\right)+C\frac{d\theta}{dX}\delta\left(\frac{d\theta}{dX}\right)\right){\rm d}X} \label{eq13}.
\end{eqnarray}
Using \eqref{eq12} and \eqref{eq13}, the expression for total internal virtual work $\delta W_{int}$ is obtained as:
\begin{align}
&\delta W_{int}=0=\int_0^L {\left [\left (h\left ( \bar{E}(\lambda-1)+ \frac{\nu}{1-\nu}\frac{\tau_s}{2}\frac{d\theta}{dX}\right )+\tau_s\right) \delta\lambda\right ]{\rm d}X}+ \nonumber \\
&\int_0^L {\left [\left (\frac{h^3}{12}\left (\bar{E}\frac{d\theta}{dX}-\frac{\nu}{1-\nu}\frac{\tau_s}{h}\frac{d\theta}{dX}  \right )-\frac{\tau_s h}{2}+C\frac{d\theta}{dX}\right )\delta \left ( \frac{d\theta}{dX} \right ) \right ] {\rm d}X}.
\label{eq16}
\end{align}
At the boundary, constraints can be applied on displacement or rotation, but not on stretch. So, we use the following relation to convert variation in stretch ($\delta \lambda$) in terms of variation in displacement and rotation (see \fref{beam_surfS}).
\begin{eqnarray}
\frac{du}{dX}=\lambda \cos{\theta}-1, \label{eq17}\\
\delta \lambda=\sec{\theta}~\delta \left ( \frac{du}{dX} \right )+\lambda \tan{\theta}~\delta\theta, \label{eq18}
\end{eqnarray}
where $u$ is displacement in $X$-direction. Replacing $\delta \lambda$ with expression in \eqref{eq18}, simplifying the terms in \eqref{eq16} using integration by parts and applying the principle of variation, we obtain the following governing equations for a beam with surface stress.
\begin{align}
\frac{d}{dX}\left[ \left( h\left ( \bar{E}(\lambda-1)+ \frac{\nu}{1-\nu}\frac{\tau_s}{2}\frac{d\theta}{dX}\right )+\tau_s\right) \sec{\theta}\right]&=0, \label{eq21}\\
\frac{d}{dX}\left [\left (\frac{h^3}{12}\left (\bar{E}\frac{d\theta}{dX}-\frac{\nu}{1-\nu}\frac{\tau_s}{h}\frac{d\theta}{dX}  \right )-\frac{\tau_s h}{2}+C\frac{d\theta}{dX}\right )\right ]&-\nonumber \\
\left( h\left ( \bar{E}(\lambda-1)+ \frac{\nu}{1-\nu}\frac{\tau_s}{2}\frac{d\theta}{dX}\right )+\tau_s \right) \lambda\tan{\theta}&=0.  \label{eq22}
\end{align}
The boundary conditions (BC) are given by the following equations:
\begin{align}
\left[ \left( h\left ( \bar{E}(\lambda-1)+ \frac{\nu}{1-\nu}\frac{\tau_s}{2}\frac{d\theta}{dX}\right )+\tau_s\right) \sec{\theta}\delta u \right]_{0,L}&=0, \label{eq23a}\\
\left [\left (\frac{h^3}{12}\left (\bar{E}\frac{d\theta}{dX}-\frac{\nu}{1-\nu}\frac{\tau_s}{h}\frac{d\theta}{dX}  \right )-\frac{\tau_s h}{2}+C\frac{d\theta}{dX}\right )\delta \theta \right ]_{0,L}&=0. \label{eq23b}
\end{align}
It should be noted that the above equations are nonlinear. Furthermore, $\tau_s$ itself depends on $\lambda$ and $\frac{d\theta}{dX}$ (see \eqref{eq4}).

The above equations can be used to determine surface stress on a cantilever beam by measuring curvature in the beam. For a cantilever beam, boundary conditions \eqref{eq23a} and \eqref{eq23b} at the free end at $X=L$ give:
\begin{align}
\left[ \left( h\left ( \bar{E}(\lambda-1)+ \frac{\nu}{1-\nu}\frac{\tau_s}{2}\frac{d\theta}{dX}\right )+\tau_s\right) \sec{\theta}\right]_{L}&=0, \label{eq23c}\\
\left [\left (\frac{h^3}{12}\left (\bar{E}\frac{d\theta}{dX}-\frac{\nu}{1-\nu}\frac{\tau_s}{h}\frac{d\theta}{dX}  \right )-\frac{\tau_s h}{2}+C\frac{d\theta}{dX}\right )\right ]_{L}&=0. \label{eq23d}
\end{align}
Using the above, governing equations \eqref{eq21} and \eqref{eq22} turn out to be:
\begin{eqnarray}
h\left ( \bar{E}(\lambda-1)+ \frac{\nu}{1-\nu}\frac{\tau_s}{2}\frac{d\theta}{dX}\right )+\tau_s=0, \label{eq24a} \\
\frac{h^3}{12}\left (\bar{E}\frac{d\theta}{dX}-\frac{\nu}{1-\nu}\frac{\tau_s}{h}\frac{d\theta}{dX}  \right )-\frac{\tau_s h}{2}+C\frac{d\theta}{dX}=0.  \label{eq24b}
\end{eqnarray}
Using (\ref{eq24b}), relation between curvature ($\kappa$) and effective surface stress ($\tau_s$) is obtained.
\begin{eqnarray}
\kappa\approx\frac{d\theta}{dX}=\frac{6\tau_s}{\bar{E}h^2-\frac{\nu}{1-\nu}\tau_s h+\frac{12C}{h}}. \label{eq25}
\end{eqnarray}
Using (\ref{eq24a}), axial stretch in the beam can be obtained.
\begin{eqnarray}
\lambda-1=-\left(1+\frac{1}{2}\frac{\nu}{1-\nu}\frac{d\theta}{dX}h\right )\frac{\tau_s}{\bar{E}h}. \label{eq26}
\end{eqnarray}
The term $\frac{1}{2}\frac{\nu}{1-\nu}\frac{d\theta}{dX}h$ appears because of Young-Laplace correction.

To obtain $\kappa$ for a known initial surface stress $\tau_s^0$, we invoke small strain assumption and linearize \eqref{eq25} using $\kappa h << 1$ (see \ref{LinearEq}) to obtain:
\begin{equation}
\boxed{
\kappa h\approx\frac{6\tau_s^0}{\bar{E}h+4E_s+\frac{12C}{h^2}+\frac{12CE_s}{\bar{E}h^3}-\frac{\nu}{1-\nu}\tau_s^0}.} \label{eq27}
\end{equation}
This is the general form of equation relating curvature with surface stress in undeformed substrate configuration. By neglecting Young-Laplace correction term $\left( \frac{\nu}{1-\nu}\tau_s^0 \right)$ in the denominator, we obtain the same expression for curvature as in \citet{chhapadia2011} for no surface stress at the bottom face of the beam.\footnote{The coefficient of $E_s$ in the denominator of (60) in \citet{chhapadia2011} should be 4 (instead of 2) as in \eqref{eq28}).}
\begin{equation}
\kappa h \approx \frac{6\tau_s^0}{\bar{E}h+4E_s+\frac{12C}{h^2}+\frac{12CE_s}{\bar{E}h^3}}. \label{eq28}
\end{equation}
For this case, neutral axis is independent of surface stress and Young's modulus of the surface layer and is given by:
\begin{equation}
Z_n=-\frac{h}{6}-\frac{2C}{\bar{E}h^2}. \label{eq28b}
\end{equation}
On discarding curvature elasticity contribution as well by substituting $C = 0$, we obtain a curvature relation that can also be obtained from equations in \citet{begley2005} for a cantilever beam.
\begin{equation}
\kappa h \approx \frac{6\tau_s^0}{\bar{E}h+4E_s}\approx \frac{6\frac{\tau_s^0}{\bar{E}h}}{1+4\frac{E_s}{\bar{E}h}}. \label{eq29}
\end{equation}
The above relation suggests a pertinent length scale as a ratio of elastic modulus of surface and that of the substrate at which effect of surface elastic modulus becomes significant. In \citet{chhapadia2011}, molecular dynamics simulation of silver nanowire gives $E_s\sim10~N/m$. For silver, $\bar{E}\sim10^{11}~Pa$. Hence, surface elastic modulus becomes significant for substrate thickness $\sim 10^{-10}~m$.

It is shown in Section~\ref{PB_surf} that $E_s\sim|\tau_s^0|$ for polymer brush. Now, $\kappa h<<1$ implies $\tau_s^0/(\bar{E}h)<<1$ and hence $E_s/(\bar{E}h)<<1$. So the term $E_s/(\bar{E}h)$ in denominator can be neglected to obtain the famous Stoney's relation.
\begin{equation}
\kappa h \approx \frac{6\tau_s^0}{\bar{E}h}. \label{eq29b}
\end{equation}

%%%%%%%%%%%%%%%%%%%%%%%%%%%%%%%%%%%%%%%%%
\section{Experiments}
\label{exp}
We experimentally estimate the surface stress due to polymer brush along with estimation of graft density of brush and molecular weight of polymer chains in the brush. To estimate surface stress due to polymer brush, a temperature sensitive random copolymer brush, Poly(N- isopropylacrylamide)-co-Poly(N,N-Dimethylacrylamide) (PNIPAm-co-PDMA) brush, was grafted on one side of a plasticized poly(vinyl chloride) (pPVC) film using surface-initiated atom transfer radical polymerization (SI-ATRP) method \citep{matyjaszewski2001,zou2011}. In polymerization solution, $90\%$ of monomer by weight is NIPAm as against $10\%$ DMA. The polymerization time controls graft density and molecular weight of the polymer chains in the brush. This method is capable of producing high graft density brushes with large molecular weights of the polymer chains in the brush leading to large residual surface stress. Also, since the brush is sensitive to temperature, the change in surface stress with temperature is also estimated. Samples for the experiment were prepared with polymerization times of 16, 14, 12 and 9 hours, which gave brushes with decreasing graft density.

\begin{figure}[!h]
	\center
	\includegraphics[width=6.9cm]{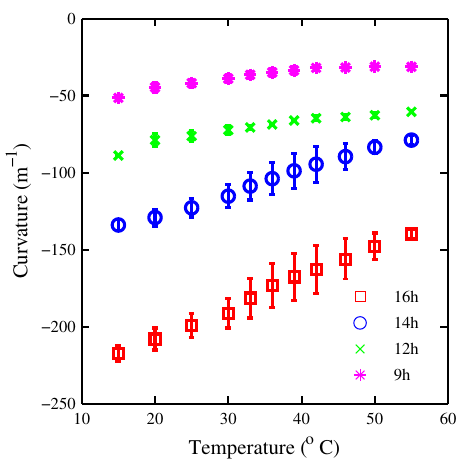}
	\hspace{-0.5cm}
	\includegraphics[width=6.9cm]{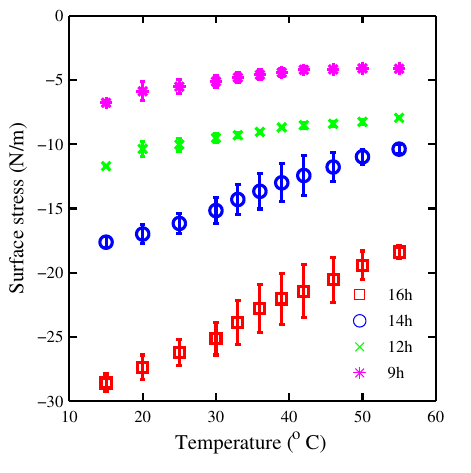}
	\caption{Variation of curvature of brush coated beams (left) and the surface stress (right) with temperature for polymer brushes with different polymerization times. Increasing temperature leads to decrease in magnitude of surface stress. Also, magnitude of surface stress  decreases with decreasing polymerization time.}
	\label{Curvature_T_exp}
\end{figure}

The pPVC beams used in the experiment are $2~cm$ long (apart from $0.5~cm$ length clamped between glass slides), $5~mm$ wide, and $254~\mu m$ thick. Young's modulus  for the pPVC film is measured to be $11.13~MPa$. We take a typical value of Poisson's ratio for pPVC of $0.4$. A brush coated beam assumes a curved shape in water.
\begin{figure}[!h]
	\centering
	\begin{subfigure}[b]{0.65\textwidth}
		\includegraphics[width=0.92\textwidth]{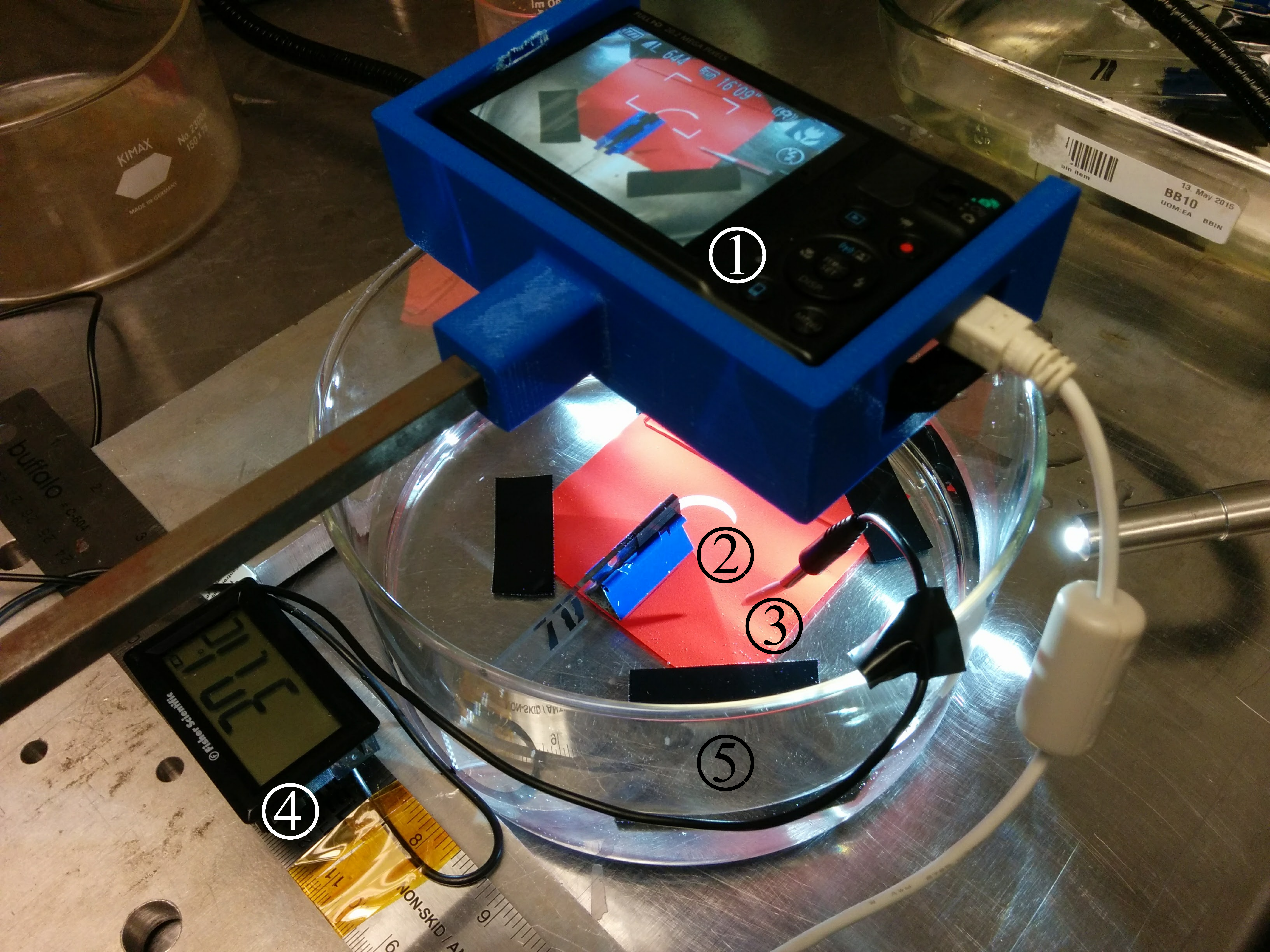}
		\caption{}
	\end{subfigure}
	\begin{subfigure}[b]{0.3\textwidth}
	\includegraphics[width=0.98\textwidth]{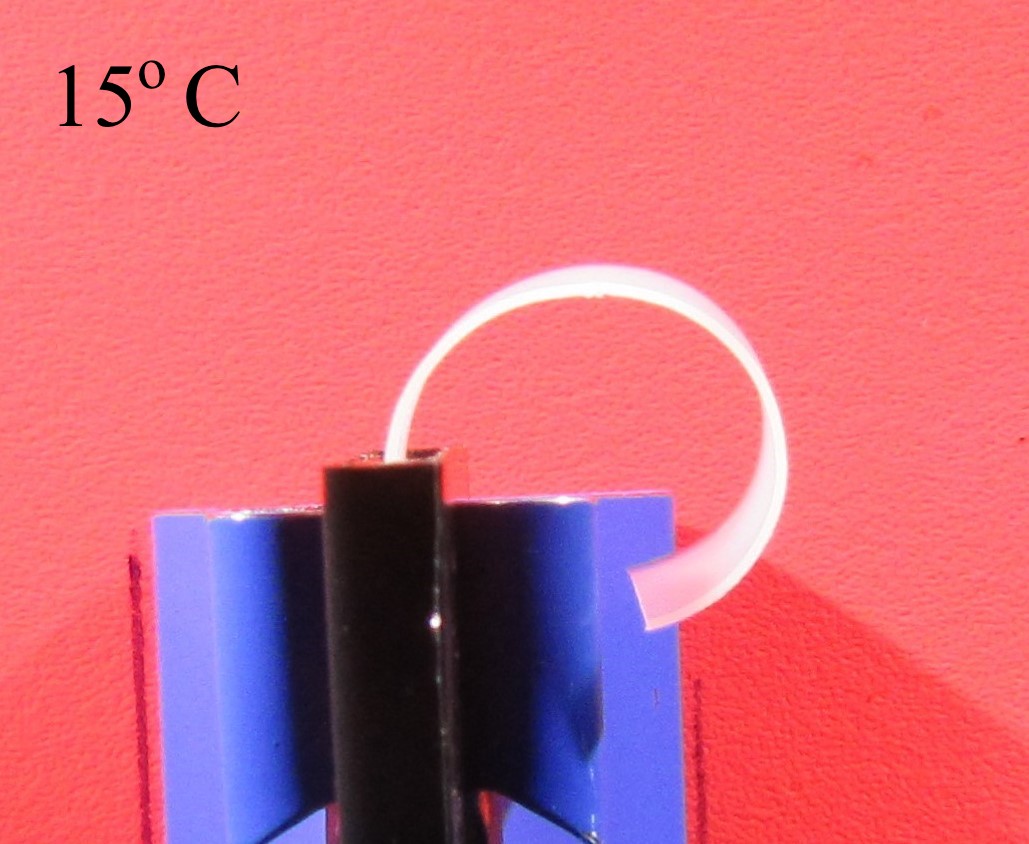} \\
	%\hspace{5mm}
	\includegraphics[width=0.98\textwidth]{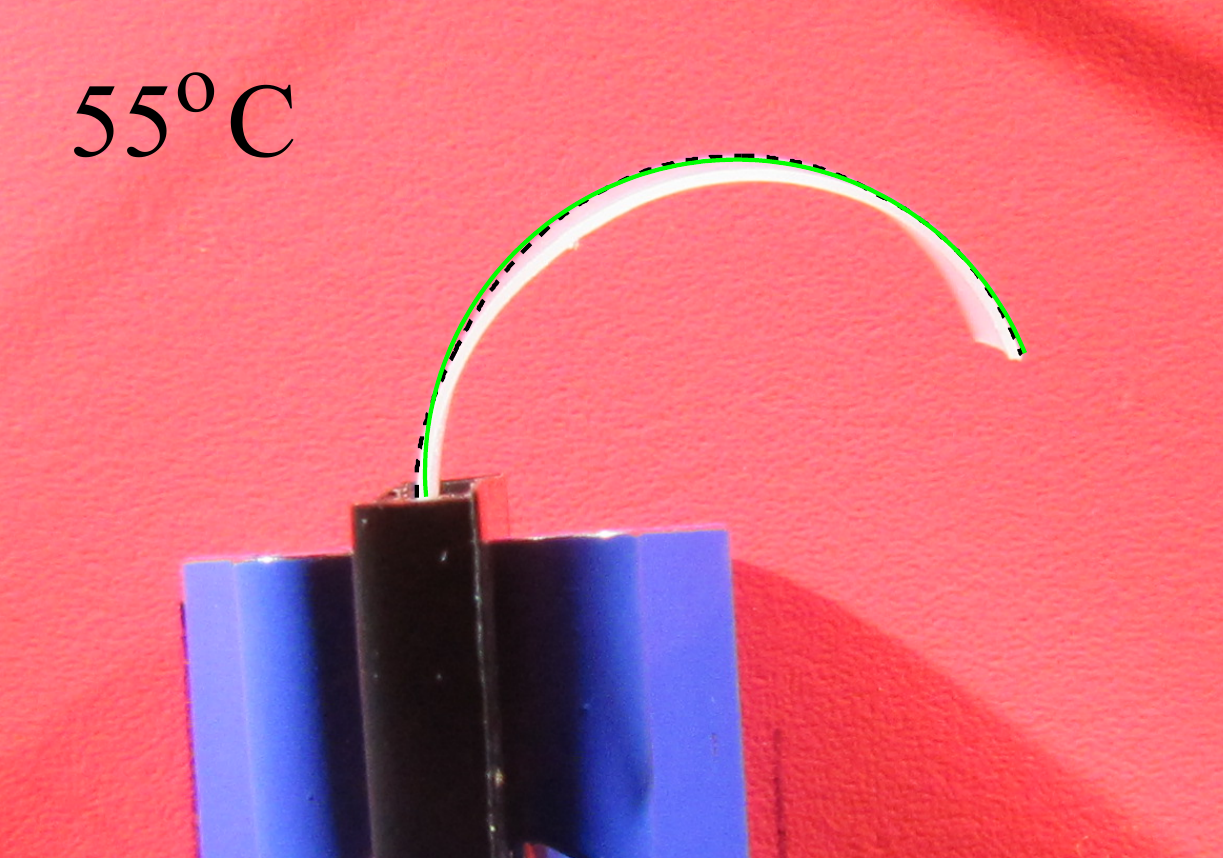}
	\caption{}
	\end{subfigure}
	~ 
	\caption{(a) Experimental set-up to measure curvature of a cantilever beam (1. camera, 2. brush coated beam, 3,4. temperature sensor and display, and 5. temperature bath), and (b) brush coated beam at $15~^\circ C$ (top) and $55~^\circ C$ (bottom). Experimentally obtained image of a cantilever beam with edge (black dashed line) traced using image processing in MATLAB and the arc of the circle fitted to the traced edge (green line).}
	\label{coated_beam}
\end{figure}
The curvature of the beam changes as temperature is varied, due to a change in surface stress (see Fig. \ref{coated_beam}(b)). To measure curvature of a beam coated with the brush at different temperatures, a water bath (see Fig. \ref{coated_beam}(a)) was brought to the desired temperature and then the beam was placed in it. Temperature of the bath is varied from $15~^\circ C$ to $55~^\circ C$ by adding hot water to the bath. Note that the same level of water in the bath is maintained throughout the experiment by taking out requisite amount of water after mixing added hot water. After placing the beam in the bath at the desired temperature, we waited for two minutes for the brush to reach steady state before taking a photograph of the beam. During this waiting time, small amount of hot water was added to the bath to ensure that temperature stays within $\pm 0.5~^\circ C$ of the desired temperature level. The beam was photographed at different temperatures by a fixed camera. Using image processing in MATLAB, the edge of the cantilever beam was traced, and using circle fitting, curvature of the coated beam was determined (see \fref{coated_beam}(b)). Note that some inaccuracy in measurement is introduced by the fact that the traced edge is not the same edge of the cantilever throughout its length. Near the fixed end of the beam, bottom edge of the beam is traced but near the free end, top edge of the beam is traced. \fref{Curvature_T_exp} shows the variation of curvature of the beam with temperature.

From \fref{Curvature_T_exp}, magnitude of curvature is ($|\kappa |$) $\sim 10^2~m^{-1}$ and the thickness of the substrates is $h$ $\sim 10^2~\mu m$. Hence the magnitude of strains ($\sim|\kappa h|$) are $\sim 10^{-2}$. Therefore the assumption of small strain in the beams in the experiment is valid. So, we can use the measured curvature, $\kappa$, to find surface stress $\tau_s$ in the beam using expression in \eqref{eq25} with the added assumption that $C\approx 0$, since substrate thickness is much higher than the height of brush ($\sim 1~\mu m$).
\begin{equation}
\tau_s=\frac{\bar{E}h^2\kappa}{6+\kappa h+\frac{\nu}{1-\nu}\kappa h}.  \label{eq48}
\end{equation}
Note that we have discarded curvature elasticity contribution in (\ref{eq48}) because our beam is much thicker than the polymer brush layer. The surface stress variation with temperature is as shown in \fref{Curvature_T_exp}. The maximum correction in the effective surface stress due to the Young-Laplace term in the denominator ($\frac{\nu}{1-\nu}h\kappa$) in (\ref{eq48}) is $<1\%$, suggesting that the term can be discarded.

It can be observed from \fref{Curvature_T_exp} that compressive surface stress due to brush decreases in magnitude with decreasing polymerization time. Also, magnitude of the surface stress can be decreased by $\approx 10~N/m$ by changing temperature from $15~^\circ C$ to $55~^\circ C$. The change in temperature induces conformation change in a thermoresponsive polymer solution at the lower critical solution temperature (LCST) of the polymer. However, the change in surface stress observed in our experiment is not sharp and occurs gradually over a range of temperature near LCST of PNIPAm ($=32~^\circ C$). It has been shown in \citet{zhulina91} that the transition in classical polymer\footnote{For a classical polymer, unlike PNIPAm, increasing temperature improves solubility of polymer in solvent}  brush is affected by surface morphology and is not a phase transition but a cooperative conformational transition for a planar brush. So, the surface morphology may be causing the observed gradual transition. This may also result from vertical phase separation in PNIPAm brush instead of a complete conformation change \citep{halperin2011,varma2016}. Presence of PDMA, which is hydrophilic at temperatures near $15 - 55~^\circ C$, may also contribute to the gradual transition as well as nonzero surface stress at heigh temperature. A complete explanation is a work for the future.

We also estimated molecular weight of polymer chains and graft density of the brush. The number averaged molecular weight ($M_n$) of the polymer chains was estimated by gel permeation chromatography (GPC) (see~\citet{zou2011} for details) to be $\approx 10^6~g/mol$. The degree of polymerization (number of monomers in a polymer chain) in the brush is $N=M_n/113.16\approx 10^4$ as molecular weight of NIPAm is $113.16~g/mol$ and we neglect contribution of PDMA to molecular weight. Graft density of the brush was estimated from dry thickness of brush (measured using SEM) (see~\citet{zou2011} for details) and was found to be $\approx 0.5~chains/nm^2$ for 14 hours of polymerization. In the absence of accurate measurement of Kuhn length of PNIPAm in literature \citep{halperin2011}, we assume effective monomer size $a\approx 1~nm$. We take the binary interaction parameter for PNIPAm to be $v=2a^3\times(31.4\times(1-T/307.80))$, where $T$ is temperature in Kelvin, as given in \citet{halperin2011} based on Flory-Huggins parameter reported in~\citet{afroze2000}. At $T=20~^{\circ}C$, surface stress is estimated to be $\approx -20~N/m$ using \eqref{eqs2}. We see that stress predicted is qualitatively of the same order of magnitude as measured directly from curvature in~\fref{Curvature_T_exp}. 

%%%%%%%%%%%%%%%%%%%%%%%%%%%%%%%%%%%%%%%%%%%

\section{Conclusion}
\label{conclusion}
This work is aimed at understanding stress in a polymer brush and the mechanics of a soft beam with polymer brush grafted on one of its surfaces. Main conclusions are as follows.
\begin{enumerate}
\item We have derived the distribution of lateral stress in polymer brush grafted to a rigid substrate using mean field theory and shown that the stress variation is quartic along height direction of brush in a good solvent as given by \eqref{eqs24}. Maximum stress occurs near the grafting surface and the stress smoothly goes to zero at the free surface of the brush.
\item The expression for stress in $\theta$-solvent is also derived and it shows a variation different from the good solvent case, and a stronger dependence on graft density as seen in \eqref{eqs24f}.
\item In the expression for stress in brush, nonbonded interaction contributes twice as much as the chain stretching in a good solvent as observed in \eqref{eqs24}. Their contributions are equal in a $\theta$-solvent.
\item Surface stress due to polymer brush has stronger dependence on graft density than on molecular weight of polymer chains in the brush as suggested by \eqref{eqs2}. 
\item Surface stress and the surface elastic modulus for a brush layer are of the same order of magnitude. Hence, the effect of surface elasticity in small strain deformation is negligible (see \eqref{eqs2}, \eqref{eqs2a}).
\item Governing equations for finite deformation (but small strain) of a beam with a coating of polymer brush on its top surface have been derived \eqref{eq27}. The Young-Laplace term, needed to satisfy equilibrium of surface layer, and effect of curvature on elasticity of surface layer have also been  included in the equations. On neglecting the contributions of the Young-Laplace effect and the curvature elasticity of surface, the expression for curvature of a beam with surface layer reduces to the results reported in literature.
\item The curvature - surface stress relation obtained using the continuum mechanics model has been used to estimate surface stress due to a thermoresponsive polymer brush grafted on a pPVC beam experimentally. The surface stress due to the brushes is found to be of the order of $-10~N/m$ and it can be decreased in magnitude by $\approx 10~N/m$ by increasing temperature from $15~^\circ C$ to $55~^\circ C$.
\item Graft density and molecular weight of the brush have also been estimated experimentally and used to estimate surface stress. The estimated surface stress is of the same order of magnitude, however no quantitative conclusion can be drawn due to large number of approximations involved in the estimation.
\end{enumerate}

Quartic variation of stress in a brush has been confirmed by molecular dynamics simulation and will be reported as a separate study. Further extension of the theory proposed is to include the effect of temperature on PNIPAm-co-PDMA brush by incorporating temperature dependent polymer-solvent interactions. Furthermore, the analytical expressions for stress distribution in brush and the surface stress and surface elasticity of brush are based on mean field theory for moderate graft density. It can be extended to higher graft density in future using numerical studies utilizing molecular dynamics simulations.

\section*{Acknowledgement}
We thank Dr. Madhab Bajgai for his contribution in the development of protocol for fabrication of polymer brush, and Madeshwaran Selvaraj and Diana Nino for their help with experimental set-up. We are grateful to Dr. J. N. Kizhakkedathu for allowing use of his lab for brush fabrication. Funding support from the Canadian Institute of Health Research (CIHR) and the Natural Sciences and Engineering Research Council of Canada (NSERC) is gratefully acknowledged.

%\section*{References}
\bibliography{references}

\appendix

\section{Comparison of brush heights obtained from different theories}
\label{brushHeight}
Here we obtain the variation in height of a polymer brush with a change in solvent quality using mean field Flory theory, and compare it with SST prediction in \citet{zhulina91} and scaling theory based prediction in \citet{utz08}.

To start with, we assume a step profile for the brush. Then, monomer density in the brush is a constant and is given by $N\rho_g/H$. Employing Flory like arguments, we can write the free energy of a polymer chain in a brush as a sum of the contributions from the chain stretching and the polymer solvent interaction, accounting for ternary interaction as follows:
\begin{equation}
F_{step}=\left(\frac{3H^2}{2Na^2}+\frac{1}{2}v\left(\frac{N\rho_g}{H}\right)^2\frac{H}{\rho_g}+\frac{1}{6}w\left(\frac{N\rho_g}{H}\right)^3\frac{H}{\rho_g}\right)k_bT,
\end{equation}
where $H/\rho_g$ is the volume occupied by a polymer chain. By minimizing $F_{step}$ with respect to the height of the brush ($H$) ($dF_{step}/dH=0$), we can obtain the expression for the height of the brush. Below we discuss height approximation in good, $\theta$ and poor solvent regime.

In a good solvent condition, binary interaction dominates and contribution from ternary interaction can be ignored \citep{zhulina91}. In $\theta$ solvent, $v=0$ and hence contribution from binary interaction is zero. In a poor solvent with large magnitude of binary interaction parameter but with a negative sign, energy contribution from the chain stretching can be ignored \citep{zhulina91}. The above assumptions allow derivation of asymptotic expressions for the height of a brush in solvents of different solvent qualities. The tables below show the heights obtained from the above approximation with predictions from SST and scaling theory.
\begin{table}[!h]
	\caption{\label{height_comparison} Comparison of expressions for height of a brush in good (though applicability is limited to solvent conditions leading to weak excluded volume interaction), $\theta$ and poor solvents obtained from mean field Flory theory~\citep{alexander77} and SST \citep{zhulina91} and scaling theory~\citep{de80,utz08}.}
	\center
	\resizebox{\linewidth}{!}{
	\begin{tabular}{|p{2.34cm}| p{3.8cm}| p{4.15cm}| p{3.5cm}|}
		\hline
		 & Flory arguments & SST	 &Scaling\\ \hline
		Good solvent & $\left(\frac{1}{6}\right)^{1/3}v^{1/3}\rho_g^{1/3} a^{2/3}N$ & $\left(\frac{4}{\pi^2}\right)^{1/3}v^{1/3}\rho_g^{1/3} a^{2/3}N$ &$\sim\rho_g^{1/3}a^{5/3}N$\\ \hline
		$\theta$ solvent & $\left(\frac{1}{9}\right)^{1/4}w^{1/4}\rho_g^{1/2} a^{1/2}N$ & $\frac{4}{\pi}\left(\frac{1}{12}\right)^{1/4}w^{1/4}\rho_g^{1/2} a^{1/2}N$ & $\sim\rho_g^{1/2}a^2N$\\ \hline
		Poor solvent & $\frac{2}{3}\frac{w}{|v|}\rho_g N$ & $\frac{2}{3}\frac{w}{|v|}\rho_g N$&  $\sim a^3\rho_g |1-2\chi|^{-1} N$\\ \hline
	\end{tabular}
	}
\end{table}

Scaling of brush height with $N$ and $\rho_g$ obtained from each of the theories is the same. Also, for a classical polymer, $v=a^3(1-2\chi)$ and $w=a^6$ \citep{de79}. So, we can see that scaling with respect to monomer size is also consistent.

For the good solvent case, SST gives $34\%$ larger brush height compared to the height derived from Flory arguments, but it is $26\%$ lower than the prediction from the blob model.

%***************************************************************************************************
\section{Linearization of cantilever beam equation}
\label{LinearEq}
Let us first define a nondimensionalization scheme.
\begin{eqnarray}
\tilde{\tau}_s=\frac{\tau_s}{{\bar{E}h}}, \qquad \tilde{\tau}_s^0=\frac{\tau_s^0}{{\bar{E}h}}, \qquad \tilde{C}=\frac{C}{{\bar{E}h^3}}, \qquad \tilde{E}_s=\frac{E_s}{{\bar{E}h}} \nonumber \\
\tilde{\kappa}=\kappa h\approx \frac{d\beta}{dX}h. \label{eq30}
\end{eqnarray}

Under this scheme, \eqref{eq24a} and \eqref{eq24b} transform to:
\begin{eqnarray}
(\lambda-1)+ \frac{1}{2}\frac{\nu}{1-\nu}\tilde{\tau}_s\tilde{\kappa}+\tilde{\tau}_s=0 \label{eq31a} \\
\frac{1}{12}\left (\tilde{\kappa}-\frac{\nu}{1-\nu}\tilde{\tau}_s\tilde{\kappa}  \right )-\frac{1}{2}\tilde{\tau}_s+\tilde{C}\tilde{\kappa}=0.  \label{eq31b}
\end{eqnarray}
\eqref{eq31a} is used to obtain stretch in the mid plane.
\begin{eqnarray}
(\lambda-1)=-\left(1+\frac{1}{2}\frac{\nu}{1-\nu}\tilde{\kappa}\right)\tilde{\tau}_s \label{eq32}
\end{eqnarray}
Also, using \eqref{eq4}, the nondimensionalized effective surface stress $\tilde{\tau}_s$ can be expressed as:
\begin{eqnarray}
\tilde{\tau}_s=\tilde{\tau}_s^0+\tilde{E}_s\left ((\lambda-1)-\frac{\tilde{\kappa}}{2}\right ). \label{eq35}
\end{eqnarray}
Substituting expression for $\lambda$ in \eqref{eq32} into the above equation and solving for $\tilde{\tau}_s$ gives:
\begin{eqnarray}
\tilde{\tau}_s=\frac{\tilde{\tau}_s^0-\tilde{E}_s\frac{\tilde{\kappa}}{2}}{1+\tilde{E}_s\left ( \frac{\nu}{1-\nu}\frac{\tilde{\kappa}}{2}+1 \right )}. \label{eq36}
\end{eqnarray}
Substituting the above expression for $\tilde{\tau}_s$ in \eqref{eq31b}, and expressing the equation as a polynomial equation in $\kappa$ gives:
\begin{eqnarray}
\frac{1}{2}\frac{\nu}{1-\nu}\tilde{E}_s\left( 2+12\tilde{C}\right)\tilde{\kappa}^2+ \nonumber \\
\left(1+4\tilde{E}_s+12\tilde{C}(1+\tilde{E}_s)-\frac{\nu}{1-\nu}\tilde{\tau}_s^0 \right)\tilde{\kappa}-6\tilde{\tau}_s^0=0. \label{eq37}
\end{eqnarray}
It can be noticed that the coefficients of $\tilde{\kappa}^2$ is of the same order or of order smaller than the coefficient of $\tilde{\kappa}$. So, for $\tilde{\kappa}<<1$, quadratic term in $\tilde{\kappa}$ can be ignored to obtain the following expression for $\tilde{\kappa}$:
\begin{eqnarray}
\tilde{\kappa}=\frac{6\tilde{\tau}_s^0}{1+4\tilde{E}_s+12\tilde{C}(1+\tilde{E}_s)-\frac{\nu}{1-\nu}\tilde{\tau}_s^0}. \label{eq38}
\end{eqnarray}
Shifting back to the dimensional form, we obtain:
\begin{equation}
\kappa h\approx\frac{6\tau_s^0}{\bar{E}h+4E_s+\frac{12C}{h^2}+\frac{12CE_s}{\bar{E}h^3}-\tau_s^0\frac{\nu}{1-\nu}}. \label{eq39}
\end{equation}

%***************************************************************
\end{document}